\documentclass[twoside]{article}
\usepackage{fleqn,espcrc2}

\usepackage{epsfig}

\newcommand{\lesssim}
{\:\lower 0.4ex\hbox{$\stackrel{\scriptstyle <}{\scriptstyle\sim}$}\:}
\newcommand{\gtrsim}
{\:\lower 0.4ex\hbox{$\stackrel{\scriptstyle >}{\scriptstyle\sim}$}\:}
\title{Neutrinos and the Supernova Origin of the Elements}

\author{Yong-Zhong Qian\address{School of Physics and Astronomy,
        University of Minnesota, Minneapolis, MN 55455, USA}}
       
\begin{document}

\begin{abstract}
Intense fluxes of neutrinos are emitted by the hot neutron star
produced in a supernova. The average supernova neutrino energies
satisfy a robust hierarchy $\langle E_{\nu_e}\rangle<
\langle E_{\bar\nu_e}\rangle<\langle E_{\nu_{\mu(\tau)}}\rangle
\approx\langle E_{\bar\nu_{\mu(\tau)}}\rangle$.
The $\nu_e$ and $\bar\nu_e$ capture reactions on neutrons and protons,
respectively, provide heating to drive a wind from the hot neutron
star. The same reactions also determine
the neutron-richness of the wind material. 
Nucleosynthesis via rapid neutron capture, the $r$-process, may occur
in the wind material as it expands away from the neutron star.
The neutron-richness of the wind material, and hence, the $r$-process
nucleosynthesis therein, are sensitive to mixing between 
$\nu_{\mu(\tau)}$/$\bar\nu_{\mu(\tau)}$ 
and $\nu_e$/$\bar\nu_e$ (or sterile neutrinos $\nu_s$/$\bar\nu_s$)
at the level of
$\sin^2 2\theta\lesssim 10^{-4}$ for $\delta m^2\gtrsim 1$~eV$^2$.
Indirect arguments and direct tests for 
the supernova origin of the $r$-process elements are discussed with
a goal to establish supernova $r$-process nucleosynthesis as an 
important probe for neutrino mixing.
\end{abstract}

\maketitle

\section{INTRODUCTION TO SUPERNOVA $r$-PROCESS NUCLEOSYNTHESIS}
A star lives a luminous life by burning H into successively heavier
elements. However, as the Fe group nuclei near mass number $A=56$ are 
most tightly bound, no more nuclear binding energy can be released to 
power the star by burning ``Fe.'' Therefore, heavy elements beyond 
``Fe'' have to be made by processes other than normal stellar burning. 
One such process is the rapid neutron capture process, or the 
$r$-process for short. This process is responsible for approximately 
half the natural abundance of nuclei with mass numbers $A>100$. 
Typical $r$-process elements are Eu, Pt, U, and Th.
A crude picture for the $r$-process is as follows. One starts with 
some seed nuclei and lots of neutrons. The seed nuclei then rapidly 
capture these neutrons to make very neutron-rich unstable progenitor 
nuclei. After neutron capture stops, the progenitor nuclei successively
$\beta$-decay towards stability and become the $r$-process nuclei
observed in nature.

The $r$-process has a lot to do with supernova neutrinos.
Neutrinos and antineutrinos of all three flavors are emitted by the
neutron star produced in a supernova. The individual neutrino species
has approximately the same luminosity but very different average energy.
As the neutrinos diffuse out of the neutron star, they thermally
decouple from the neutron star matter at different radii due to the 
difference in their ability to exchange energy with such matter.
With higher temperatures at smaller radii, $\nu_\mu$, $\nu_\tau$, and 
their antineutrinos decoupling at the smallest radii are imprinted with
the highest average energy while $\nu_e$ decoupling at the largest
radii are imprinted with the lowest average energy. The average energy
of $\bar\nu_e$ lies between those of $\nu_{\mu(\tau)}$ and $\nu_e$. 
Typical average supernova neutrino
energies are $\langle E_{\nu_e}\rangle\approx 11$~MeV,
$\langle E_{\bar\nu_e}\rangle\approx 16$~MeV, and
$\langle E_{\nu_{\mu(\tau)}}\rangle\approx 
\langle E_{\bar\nu_{\mu(\tau)}}\rangle\approx 25$~MeV.
I emphasize that while different supernova calculations give somewhat
different numerical values, there is a robust hierarchy of the average
supernova neutrino energies: $\langle E_{\nu_e}\rangle<
\langle E_{\bar\nu_e}\rangle<\langle E_{\nu_{\mu(\tau)}}\rangle$.
This hierarchy is the most crucial aspect of supernova neutrino 
emission relevant for our discussion.

A few seconds after the supernova
explosion, we have a hot neutron star near the center of the supernova.
The neutron star is still cooling by emitting neutrinos. The shock
wave which makes the supernova explosion is far away from the neutron
star. On its way out to make the explosion, the shock wave has cleared
away almost all the material above the neutron star, leaving behind
only a thin atmosphere. Close to the neutron star, the temperature is
several MeV and the atmosphere is essentially dissociated into neutrons
and protons. As the neutrinos emitted by the neutron star free-stream
through this atmosphere, some of the $\nu_e$ and $\bar\nu_e$ are
captured by the neutrons and protons and their energy is deposited
in the atmosphere. In other words, the atmosphere is heated by the
neutrinos. As a result, it expands away from the neutron star and
eventually develops into a mass outflow --- a neutrino-driven ``wind''
\cite{duncan,qwoo}.

The capture reactions $\nu_e+n\to p+e^-$ and 
$\bar\nu_e+p\to n+e^+$ not only provide heating to drive the wind,
but also interconvert neutrons and protons. With a significant excess 
of $\langle E_{\bar\nu_e}\rangle$ over $\langle E_{\nu_e}\rangle$, neutron 
production by the $\bar\nu_e$ dominates neutron destruction by the
$\nu_e$. Consequently, the wind material is neutron rich.
As this material expands away from the neutron star, its temperature
and density decrease and various nuclear reactions take place to
change its composition. When the temperature drops to $\approx 0.5$~MeV,
essentially all the protons are assembled into $\alpha$-particles and
the material at this temperature just contains neutrons and 
$\alpha$-particles. As the temperature drops further, 
$\alpha$-particles and neutrons are burned into heavier nuclei
(the $\alpha$-process \cite{wooh}).
By the time the Coulomb barrier stops all charged-particle reactions
at a temperature of $\approx 0.25$~MeV, nuclei with $A\sim 100$ have
been produced. These nuclei then become the seed nuclei to capture
the remaining neutrons during the subsequent $r$-process, which
occurs at temperatures below $\approx 0.25$~MeV (e.g., 
\cite{meyer1,janka1,woo}).

\section{NEUTRINO OSCILLATIONS AND SUPERNOVA $r$-PROCESS NUCLEOSYNTHESIS}
\begin{figure}
\vspace{5pt}
\centerline{\epsfig{file=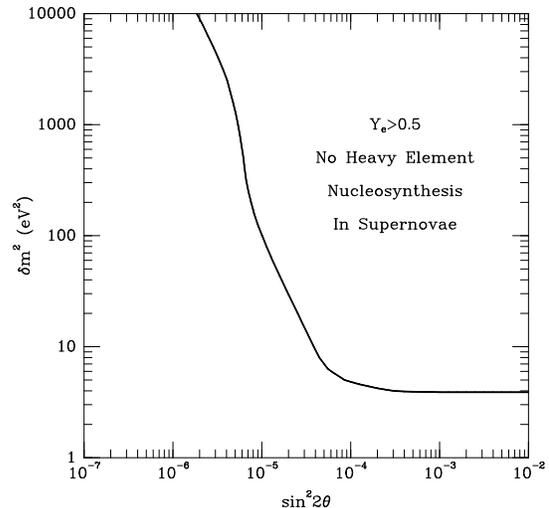,width=2.8in}}
\caption{Matter-enhanced mixing between $\nu_{\mu(\tau)}$ and $\nu_e$
with parameters in the labeled region drives the material in 
the neutrino-driven wind
proton rich (corresponding to an electron fraction $Y_e>0.5$), and hence,
is incompatible with supernova $r$-process nucleosynthesis.}
\label{mix}
\end{figure}

An absolutely necessary condition for an $r$-process to occur in
the neutrino-driven wind is that the wind material must be neutron rich.
As $\nu_{\mu(\tau)}$ have the highest average energy,
significant mixing between $\nu_{\mu(\tau)}$ and $\nu_e$ would increase
the destruction of neutrons by $\nu_e$ and drive the wind material
proton rich \cite{qetal}. 
(Even in the extreme case of $\langle E_{\nu_e}\rangle<
\langle E_{\bar\nu_e}\rangle\approx\langle E_{\nu_{\mu(\tau)}}\rangle$,
conversion of $\nu_{\mu(\tau)}$ into $\nu_e$ can still drive the wind
material proton rich as neutron production by the $\bar\nu_e$ is hindered
while neutron destruction by the $\nu_e$ is aided by the neutron-proton mass 
difference.) For $\nu_{\mu(\tau)}$ with a cosmologically
significant (vacuum) mass of $\sim 1$--100 eV, matter-enhanced mixing with 
a lighter $\nu_e$ would occur below the region where the neutron-richness
of the wind material is determined \cite{qetal}.
Therefore, such mixing is severely constrained if the 
$r$-process indeed occurs in the neutrino-driven wind in a supernova. 
Figure \ref{mix} shows the parameters for mixing between 
$\nu_{\mu(\tau)}$ and $\nu_e$ that are incompatible with supernova 
$r$-process nucleosynthesis \cite{qetal,qfull}. Note that the 
$\nu_\mu$-$\nu_e$ mixing parameters
$\sin^2 2\theta\sim 3\times 10^{-3}$--$10^{-2}$ at $\delta m^2\gtrsim$ 
several eV$^2$ reported by the LSND experiment \cite{lsnd}
lie in the incompatible region. Thus only the LSND parameters at 
$\delta m^2\lesssim$ several eV$^2$ are compatible with supernova 
$r$-process nucleosynthesis. The LSND parameters at low $\delta m^2$
are also consistent with the results from the KARMEN experiment 
\cite{karmen}.

One may ask what happens if the LSND parameters at $\delta m^2\gtrsim$
several eV$^2$ turn out to be true. In this case, supernova $r$-process 
nucleosynthesis can still occur if for example, 
a light sterile neutrino $\nu_s$ 
is introduced so that $\nu_{\mu(\tau)}$ with the highest average energy
will be turned into harmless $\nu_s$ before they can mix with
the $\nu_e$ \cite{caldwell}. Of course, one can consider many other 
possibilities of neutrino mixing that are compatible with supernova 
$r$-process nucleosynthesis. In any case, supernova $r$-process 
nucleosynthesis provides
a potential probe for mixing between $\nu_{\mu(\tau)}$/$\bar\nu_{\mu(\tau)}$ 
and $\nu_{e(s)}$/$\bar\nu_{e(s)}$
at the level of $\sin^2 2\theta\lesssim 10^{-4}$ 
for $\delta m^2\gtrsim 1$~eV$^2$.
To establish this probe on a solid basis
requires us to check the supernova laboratory
against the standards of terrestrial experiments.

A terrestrial neutrino oscillation experiment consists of a known
neutrino source and a detector with unambiguous signals for neutrino 
oscillations. In the supernova laboratory, the hot neutron star is the 
neutrino source. The most crucial aspect of this source is the robust
hierarchy of average neutrino energies. As the neutrinos move away
from the neutron star, various scenarios of neutrino mixing can occur. The
$\nu_e$ and $\bar\nu_e$ emerging from the mixing region are detected by
the capture reactions on neutrons and protons, respectively. The signal
that we are looking for is the production of $r$-process elements such as
Eu and U in supernovae. If this signal is observed, any scenario of
neutrino mixing that
would cause neutron destruction by the $\nu_e$ to dominate neutron
production by the $\bar\nu_e$ is forbidden. 

Note that the use of
supernova nucleosynthesis to study forbidden scenarios of neutrino 
mixing only relies on the {\it necessary} condition for an $r$-process 
to occur in the neutrino-driven wind --- the wind material must be 
neutron rich. Provided that we can prove the supernova origin of 
the $r$-process
elements, we do not have to understand the exact details of
supernova $r$-process nucleosynthesis
in order to use the supernova laboratory in the ``forbidden'' mode 
for studying neutrino mixing. On the other hand,
if we also know the exact characteristics of supernova neutrino emission and 
of the neutrino-driven wind, we can even determine whether 
a certain neutrino mixing scenario is required based on the {\it sufficient} 
conditions for supernova $r$-process nucleosynthesis. However, great
improvements in our understanding of supernovae have to be made before
the supernova laboratory can be used in the ``required'' mode for studying 
neutrino mixing. As the
supernova origin of the $r$-process elements is the basis for
using supernova nucleosynthesis to study neutrino mixing in either the 
forbidden or the required mode, it will be the focus of 
the following discussion.

\section{SUPERNOVA ORIGIN OF THE $r$-PROCESS ELEMENTS}
I first present indirect arguments for the supernova origin of
the $r$-process elements based on recent observations of $r$-process
elemental abundances in metal-poor stars and consideration of
Galactic chemical evolution. Two possible direct tests
are discussed next.

\subsection{Indirect arguments}
The astrophysical site for the $r$-process has to provide a large neutron 
abundance. This can be achieved by having $\bar\nu_e$ capture on protons
dominate $\nu_e$ capture on neutrons in a supernova. Alternatively,
neutron-rich material may be obtained in the merger of a neutron star
with another neutron star or a black hole. In fact, supernovae and
neutron star mergers are considered as the two leading candidate sites
for the $r$-process. An important distinction between these two sites
is the vast difference in the event rate. Massive stars that explode as
supernovae are a small fraction of all stars. The progenitor system
for neutron star mergers must have two massive stars in a binary.
Furthermore, this binary must survive the two supernova explosions that
produce the two compact objects for the eventual merger. A rather high
estimate of the neutron star merger rate in the Galaxy is
$\sim (3\times 10^4\ {\rm yr})^{-1}$ (e.g., \cite{nsm}).
This is still $\sim 10^3$ times smaller than the Galactic supernova rate.

Let us assume that supernovae are the major source for the $r$-process
and consider $r$-process enrichment of the interstellar medium (ISM).
The ejecta from each supernova is mixed with an average mass
$M_{\rm mix}\approx 3\times 10^4\,M_\odot$ of ISM (mostly H) 
swept up by the 
supernova remnant (e.g., \cite{snr}). For a supernova rate that is
proportional to the mass of gas, an average ISM in the Galaxy
is enriched by supernova ejecta at a frequency of
$\sim M_{\rm mix}(f_{\rm G}^{\rm SN}/M_{\rm gas})
\sim (10^7\ {\rm yr})^{-1}$, where the supernova rate per unit mass
of gas is estimated using $f_{\rm G}^{\rm SN}\sim (30\ {\rm yr})^{-1}$ and
$M_{\rm gas}\sim 10^{10}\,M_\odot$ for the present Galaxy. 
Consequently, an average ISM would be enriched with a 
solar $r$-process composition (denoted by the subscript ``$\odot,r$'')
by $\sim 10^3$ supernovae over 
a period of $\sim 10^{10}$~yr. This then determines the $r$-process
abundances resulting from a single supernova, e.g., 
(Eu/H)$_{\rm SN}\sim 10^{-3}$~(Eu/H)$_{\odot,r}$ with
Eu/H being the abundance ratio of Eu to H. In the spectroscopic
notation $\log\epsilon({\rm Eu})\equiv\log({\rm Eu/H})+12$, we have
$\log\epsilon_{\rm SN}({\rm Eu})\sim\log\epsilon_{\odot,r}({\rm Eu})-3
\approx -2.5$ \cite{qw,wq,qian}.

\begin{figure}
\vspace{5pt}
\centerline{\epsfig{file=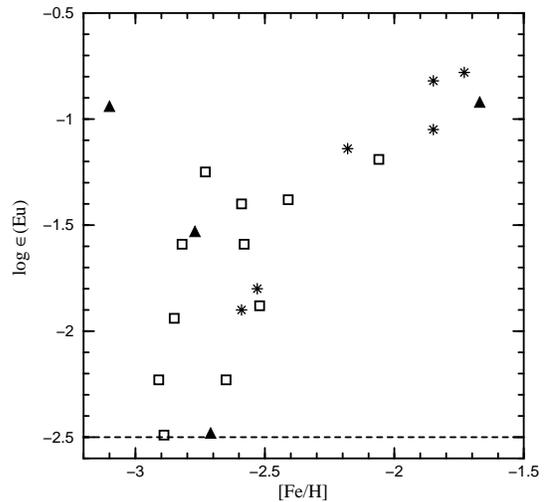,width=2.8in}}
\caption{Europium data for metal-poor stars (asterisks: 
\protect\cite{gratton}, squares: \protect\cite{mcwill},
triangles: \protect\cite{sneden1,sneden2}).
If supernovae are the major source for the $r$-process,
the Eu abundance resulting from a single event is 
$\log\epsilon({\rm Eu})\sim -2.5$ (dashed line). If neutron star
mergers were the major source for the $r$-process instead,
the Eu abundance resulting from a single event would be
$\log\epsilon({\rm Eu})\sim 0.5$.}
\label{eu}
\end{figure}

The observed Eu abundances in many metal-poor stars 
\cite{gratton,mcwill,sneden1,sneden2} are shown in 
Figure \ref{eu}. The ``metallicity'' is defined by 
[Fe/H]~$\equiv\log{\rm (Fe/H)}-\log{\rm (Fe/H)}_\odot$.
The low values of [Fe/H] for the stars
indicate that they were formed at very early times
when the ISM had been enriched by only a small number of supernovae.
The lowest Eu abundances observed in metal-poor stars
are in agreement with the $r$-process
enrichment resulting from a single supernova discussed above.

The ejecta from each neutron star merger is mixed with approximately
the same amount of ISM as swept up by a supernova remnant. However,
as the Galactic
rate of neutron star mergers is $\sim 10^3$ times smaller than that of
supernovae, an average ISM would be enriched by the ejecta from
only $\sim 1$ neutron star merger over a period of $\sim 10^{10}$~yr.
Consequently, the Eu abundance resulting from a single event would
be $\log\epsilon_{\rm NSM}({\rm Eu})\sim\log\epsilon_{\odot,r}({\rm Eu})
\approx 0.5$ if neutron star mergers were the major source for the 
$r$-process. This is in clear disagreement with the data in 
Figure \ref{eu} \cite{qian}. 

Supernovae also provided Fe enrichment of the ISM at 
[Fe/H]~$\gtrsim -2.5$ \cite{wq}. Figure \ref{eu} shows that there is
a correlation between the abundances of Eu and Fe at 
[Fe/H]~$\gtrsim -2.5$. This can be explained as the result from mixture
of the Eu and Fe produced by many supernovae if supernovae are the major
source for the $r$-process. On the other hand, 
an average ISM would be enriched in Fe by many 
supernovae between the occurrence of two successive neutron star mergers
due to the vast difference in the event rate.
Consequently, the correlation between the abundances
of Eu and Fe, especially its early onset at [Fe/H]~$\sim -2.5$,
would be very difficult to explain
if Eu enrichment were provided by neutron star mergers while Fe 
enrichment was provided by supernovae \cite{qian}.
Therefore, consideration of Galactic
chemical evolution and observations of $r$-process
elemental abundances in metal-poor stars strongly favor supernovae
over neutron star mergers as the major source for the $r$-process.
Furthermore, the total amount of $r$-process ejecta required from
each supernova to explain the observed $r$-process elemental
abundances in metal-poor stars is consistent with the amount of
material ejected in the neutrino-driven wind \cite{qwoo,qian}.
In summary, there is strong evidence for the supernova origin
of the $r$-process elements.

\subsection{Direct tests}
As described in the introduction, the $r$-process initially produces
very neutron-rich unstable progenitor nuclei. During the decay towards
stability, some progenitor nuclei decay to the excited states of their
daughters. The gamma rays from the de-excitation of the daughters 
constitute the signal for the presence of these $r$-process
progenitor nuclei. If such gamma rays are detected from a future supernova,
then we will have proven the supernova origin of the $r$-process elements. 
A supernova becomes transparent to gamma rays after 
approximately one year of expansion. Therefore, the relevant $r$-process
progenitor nuclei must have lifetimes of $\gtrsim 1$~yr. Three most
promising nuclei are $^{125}$Sb, $^{144}$Ce, and $^{194}$Os. The typical
gamma-ray flux from a supernova at a distance of 10 kpc is 
$\gtrsim 10^{-7}$~$\gamma$~cm$^{-2}$~s$^{-1}$ \cite{qvw1}. 
To detect such fluxes
requires a future supernova and a new detector with a sensitivity of
$\sim 10^{-7}$~$\gamma$~cm$^{-2}$~s$^{-1}$. Therefore, 
while this is the most
direct means to prove $r$-process production in supernovae, it is also 
the hardest.

If we can find suitable $r$-process progenitor nuclei with
lifetimes much longer than one year, we may search for the decay
gamma rays from the remnant of a past supernova. As a longer lifetime
means a smaller decay rate, to ensure a substantial gamma-ray flux
requires a nearby supernova remnant (SNR). The Vela SNR is at a distance
of $\approx 250$~pc. The age of the pulsar in this SNR is $\sim 10^4$~yr.
The relevant $r$-process progenitor nucleus
for gamma-ray detection is $^{126}$Sn with
a lifetime of $\sim 10^5$~yr and
several prominent decay gamma rays. 
The expected gamma-ray fluxes due to decay
of $^{126}$Sn in the Vela SNR
are $\gtrsim 10^{-7}$~$\gamma$~cm$^{-2}$~s$^{-1}$ \cite{qvw1}.
A new SNR near Vela was discovered recently through its X-ray emission
and the gamma rays from decay of $^{44}$Ti. As $^{44}$Ti has a lifetime
of only $\approx 90$~yr, the age of the new SNR is $\lesssim 10^3$~yr.
In this case, we can search for decay gamma rays from a number of
actinides with lifetimes of $\sim 10^3$~yr. The expected fluxes are
again $\gtrsim 10^{-7}$~$\gamma$~cm$^{-2}$~s$^{-1}$ \cite{qvw2}.

As we can see, to prove $r$-process production in supernovae by gamma-ray
astronomy requires a new detector with a sensitivity of
$\sim 10^{-7}$~$\gamma$~cm$^{-2}$~s$^{-1}$. 
By comparison, the sensitivity of
the Compton Gamma-Ray Observatory just demissioned was 
$\sim 10^{-5}$~$\gamma$~cm$^{-2}$~s$^{-1}$, and that of the INTEGRAL
experiment to be launched within the next few years 
is $\sim 10^{-6}$~$\gamma$~cm$^{-2}$~s$^{-1}$. Perhaps a sensitivity of
$\sim 10^{-7}$~$\gamma$~cm$^{-2}$~s$^{-1}$ can be reached within the next
decade.

There is yet another way to prove $r$-process production in supernovae
if we take advantage of the occurrence of supernovae in binaries. 
Approximately half of the stars are in binaries. Some
binaries initially consist of a massive star and a low-mass star.
After the massive one explodes as a supernova, it is possible for the
neutron star or black hole produced in the supernova to remain in orbit
around the low-mass star. Furthermore, a fraction of the $r$-process
ejecta from the supernova would be intercepted by the low-mass star.
Therefore, $r$-process production in supernovae will be proven if
we detect $r$-process abundance anomalies on the surface of the
binary companion to a neutron star or black hole \cite{qian}. 
Large overabundances of supernova
products such as O, Mg, Si, and S have been observed recently in
the binary companion to a black hole \cite{israel}. With the use of the
Hubble Space Telescope and the Keck Observatory, perhaps this kind
of observation can be extended successfully to the $r$-process elements 
within the next few years.

\section{CONCLUSIONS}
I have discussed the role of neutrinos in heavy element production in
supernovae, especially the effects of neutrino mixing on supernova
$r$-process nucleosynthesis. The neutron-richness of the material in
the neutrino-driven wind in a supernova is sensitive to
mixing between $\nu_{\mu(\tau)}$/$\bar\nu_{\mu(\tau)}$ and 
$\nu_{e(s)}$/$\bar\nu_{e(s)}$ at the level of
$\sin^2 2\theta\lesssim 10^{-4}$ for $\delta m^2\gtrsim 1$~eV$^2$.
A necessary condition for an $r$-process to occur in the wind is
that the wind material must be neutron rich. Provided that the supernova
origin of the $r$-process elements can be proven, this necessary 
condition can be used to eliminate any scenario 
of neutrino mixing that would cause neutron destruction by the $\nu_e$
to dominate neutron production by the $\bar\nu_e$.

I have presented indirect arguments for the supernova origin of 
the $r$-process elements based on recent observations of 
$r$-process elemental
abundances in metal-poor stars and consideration of Galactic chemical
evolution. I have also discussed two direct tests for
$r$-process production in supernovae: detection of gamma rays due to
decay of $r$-process progenitor nuclei from a future supernova or
nearby supernova remnant and observation of $r$-process abundance
anomalies on the surface of the binary companion to a neutron star or black
hole. Hopefully, these tests will prove the supernova origin of the
$r$-process elements in the near future, thereby establishing supernova
$r$-process nucleosynthesis as an extremely
sensitive probe for neutrino mixing. 

\section*{ACKNOWLEDGMENTS}
This work was supported in part by the Department of Energy under grant
DE-FG02-87ER40328.

\end{document}